\documentclass[aps,pra,showpacs,amsmath,amssymb,preprintnumbers,superscriptaddress,10pt,twocolumn,longbibliography]{revtex4-1}

\usepackage{bm}
\usepackage{graphicx}
\usepackage{SIunits}
\usepackage{hyphenat}
\usepackage[bookmarks=false]{hyperref}
\usepackage{color}
\usepackage[capitalise]{cleveref}
\crefname{section}{Sec.}{Secs.}
\Crefname{section}{Section}{Sections}
\usepackage{makecell}

\definecolor{pink}{RGB}{255,0,255}
\definecolor{pink}{RGB}{255,0,255}
\definecolor{darkergreen}{RGB}{0,180,0}

\begin{document}

\title{A low-noise single-photon detector for long-distance free-space\\quantum communication}

\author{Elena~Anisimova}
\email{anisimovaa@gmail.com}
\affiliation{Russian Quantum Center, Skolkovo, Moscow 121205, Russia}
\affiliation{NTI Center for Quantum Communications, National University of Science and Technology MISiS, Moscow 119049, Russia}
\affiliation{Institute for Quantum Computing, University of Waterloo, Waterloo, ON, N2L3G1 Canada}
\affiliation{Department of Physics and Astronomy, University of Waterloo, Waterloo, ON, N2L~3G1 Canada}

\author{Dmitri~Nikulov}
\affiliation{Institute for Quantum Computing, University of Waterloo, Waterloo, ON, N2L3G1 Canada}

\author{Simeng~Simone~Hu}
\affiliation{Institute for Quantum Computing, University of Waterloo, Waterloo, ON, N2L3G1 Canada}

\author{Mark~Bourgon}
\affiliation{Institute for Quantum Computing, University of Waterloo, Waterloo, ON, N2L3G1 Canada}

\author{Sebastian~Philipp~Neumann}
\affiliation{Institute for Quantum Optics and Quantum Information (IQOQI), Austrian Academy of Sciences, Boltzmanngasse~3, A-1090 Vienna, Austria}

\author{Rupert~Ursin}
\affiliation{Institute for Quantum Optics and Quantum Information (IQOQI), Austrian Academy of Sciences, Boltzmanngasse~3, A-1090 Vienna, Austria}

\author{Thomas~Jennewein}
\affiliation{Institute for Quantum Computing, University of Waterloo, Waterloo, ON, N2L3G1 Canada}
\affiliation{Department of Physics and Astronomy, University of Waterloo, Waterloo, ON, N2L~3G1 Canada}

\author{Vadim~Makarov}
\affiliation{Russian Quantum Center, Skolkovo, Moscow 121205, Russia}
\affiliation{\mbox{Shanghai Branch, National Laboratory for Physical Sciences at Microscale and CAS Center for Excellence in} \mbox{Quantum Information, University of Science and Technology of China, Shanghai 201315, People's Republic of China}}
\affiliation{NTI Center for Quantum Communications, National University of Science and Technology MISiS, Moscow 119049, Russia}
\affiliation{Department of Physics and Astronomy, University of Waterloo, Waterloo, ON, N2L~3G1 Canada}

\date{\today}

\begin{abstract}
We build and test a single-photon detector based on a Si avalanche photodiode Excelitas 30902SH thermoelectrically cooled to $-100~\celsius$. Our detector has dark count rate below $1~\hertz$, $500~\micro\meter$ diameter photosensitive area, photon detection efficiency around $50\%$, afterpulsing less than $0.35\%$, and timing jitter under $1~\nano\second$. These characteristics make it suitable for long-distance free-space quantum communication links, which we briefly discuss. We also report an improved method that we call long-time afterpulsing analysis, used to determine and visualise long trap lifetimes at different temperatures.
\end{abstract}

\maketitle

\section{Introduction}
\label{sec:intro}

Quantum key distribution (QKD) is the most commercialized area of quantum communication. A necessity of highly secure communications that will be able to withstand hacking attacks from quantum computers has led to fast development of quantum cryptography \cite{bennett1984, ekert1991, gilbert2000, buttler2000, rarity2002, nordholt2002}. Several companies (e.g.,\ ID~Quantique in Switzerland) offer QKD systems, which are ready for use by customers with high-demand of security, e.g.,\ in banking, medicine, government and military. The next step of QKD development is its expansion on global scale and creation of world-wide QKD network \cite{ursin2009, bonato2009, bourgoin2013, meyer-scott2011, higgins2012, gibney2016, wang2020, wei2020, chen2020, li2019}. A lot of work has been already done to reach the longest distances for free-space quantum communications \cite{buttler1998, hughes2002, ursin2007, ma2012, gibney2016, joshi2018, yin2017, liao2018, yang2019, han2020, shen2021} . 

The main challenge of long-distance free-space quantum communication are the high photon losses in the channel, caused mostly by absorption, diffraction and turbulence in the air \cite{erven2012, bourgoin2013}. To minimize absorption losses, while still using relatively simple photon detectors, a wavelength within a low-loss window at around $800~\nano\meter$ is often chosen as a good optimum. Diffraction losses can be minimized only by increasing sizes of sending and receiving telescopes, however the atmospheric turbulence puts a limit on gains of this approach. Other photon losses occur in sending and receiving systems, of which single-photon detectors are an essential part. The ideal detectors for long-distance free-space quantum communications must have high detection efficiency, low dark count rate (DCR), a sufficiently large photosensitive area, low detection timing jitter, low afterpulsing probability, compact package, low power consumption (if used in space) and cost. From a number of potential candidates, silicon avalanche photodiodes (APDs) and photomultiplier tubes (PMTs) are the most suitable. However PMTs have lower detection efficiency at the required $800~\nano\meter$, whereas APDs have a long history of use in quantum communications.

Silicon APDs (Si-APDs) have been used to detect photons of around $800~\nano\meter$ wavelength in a number of long-distance quantum experiments, on ground and in space \cite{buttler1998,hughes2002,ursin2007,ma2012,yin2017,liao2017b,ren2017,liao2018,xu2019}. They possess all the qualities required, however their DCR and afterpulsing probability vary greatly with their operating temperature and size of the photosensitive area. Dark counts are false counts produced by the photon detector in the absence of incident light. In APDs, they are caused by intrinsic processes \cite{haitz1965,cova1996,cova2004}. In silicon APDs, the main contribution to the dark counts are thermal excitations of carriers that subsequently trigger avalanches. Their rate increases exponentially with temperature \cite{lacaita1994}, thus to decrease DCR, APDs are cooled down. Other effects contributing to DCR are carrier tunneling and afterpulsing \cite{haitz1965,cova1996,cova2004}.

Dark counts contribute random errors in the communication channel. If the error rate exceeds a certain threshold, a quantum communication protocol fails \cite{bennett1992b}. The dark count rate thus needs to be kept below a certain level, which in turn depends on the rate of detection of signal photons and their registration scheme (which may involve post-selection in a time window to improve the signal-to-noise ratio). Generally, if the photon detection rate is high, a higher DCR can be tolerated. For example, in a space-based photon pair generation experiment \cite{tang2016} that did not transmit photons to the ground but detected coincidences locally in the satellite, the photon detection rate in a single detector and its dark count rate were both about $5\times10^4$ counts per second (cps). The APDs in this experiment did not require cooling and operated at about room temperature. However, transmission through a long free-space channel incurs a high loss owing to beam spreading and other factors, lowering the signal count rate greatly. In a typical satellite-to-ground link, the loss is more than $30~\deci\bel$ and the signal count rate is less than $6\times10^3$~cps \cite{liao2017b}. The detector's DCR in this experiment was under $25$~cps, requiring cooling the APDs. More challenging future applications such as communication with a geostationary orbit satellite incur a signal loss of the order of $50~\deci\bel$ (assuming the ground telescope under $1~\meter$ diameter) \cite{dirks2021} or more \cite{gunthner2017}. This requires pushing the limits of APD performance even further. Here we explore APD performance in low-temperature regime from $0$ to $-100~\celsius$ and study its DCR and afterpulsing. The latter can also be a problem in the low-temperature regime. Afterpulses are noise counts caused by spontaneously released carriers trapped during a previous avalanche, resulting in a short-time increase of DCR after each count \cite{haitz1965,cova1996,cova2004}. The lower the APD temperature, the longer the afterpulsing time is and the more likely the afterpulses introduce errors in the quantum protocol.

We have built and tested a Si-APD detector assembly that achieves very good parameters, suitable for long-distance free-space quantum communication experiments. Our SPD uses an off-the-shelf APD (Excelitas C30902SH) with $500~\micro\meter$ diameter photosensitive area. Our new detector has a compact package (see \cref{fig:mech_design,fig:detector_open}), but is able to cool down an APD down to $-100~\celsius$, utilizing a 5-stage thermoelectric cooler. Owing to this operating temperature, our detector has very low DCR, down to few counts per second. The detector package is vacuumed, to improve thermal insulation and prevent condensation. We measure its DCR, photon detection efficiency, timing jitter and afterpulsing as functions of temperature and APD bias voltage.  

For the analysis of afterpulsing we have developed a calculation method suited especially for afterpulses with long decay time, which appear in APDs at low temperatures and feature high probabilities and longer lifetimes of traps. Unlike previously described methods \cite{yoshizawa2002b, giudice2003, itzler2012, humer2015}, we analyze time intervals of SPD's outcoming pulses not only between two adjacent pulses, but between all pulses in an arbitrary chosen time window. Our method allows to measure afterpulses with lifetimes longer than the time between neighboring pulses. 

The paper is structured as follows. In \cref{sec:design} we describe our detector design. In \cref{sec:characterization} we explain our measurement protocols, and in \cref{sec:afterpulsing} we present our improved method for afterpulsing analysis. In \cref{sec:results} we demonstrate characteristics of our detector. We model the performance of a high-loss quantum key distribution experiment with a geostationary-orbit satellite in \cref{sec:modeling} and conclude in \cref{sec:discussion}.

\section{Detector design}
\label{sec:design}

\subsection{Mechanical and thermal}

Our present detector model is an improved version of the previous home-built SPD \cite{kim2011}, which was able to cool down to around $-65$ to $-80~\celsius$ and demonstrated DCR of about 20~cps that made it possible to use it in a long-distance free-space experiment \cite{ma2012}. In our present work we attempt to create an APD based SPD able to cool down below $-100~\celsius$ in a relatively compact and cheap package, and investigate behavior of Si APDs at such low temperatures.  

\begin{figure}
  \includegraphics[width=\columnwidth]{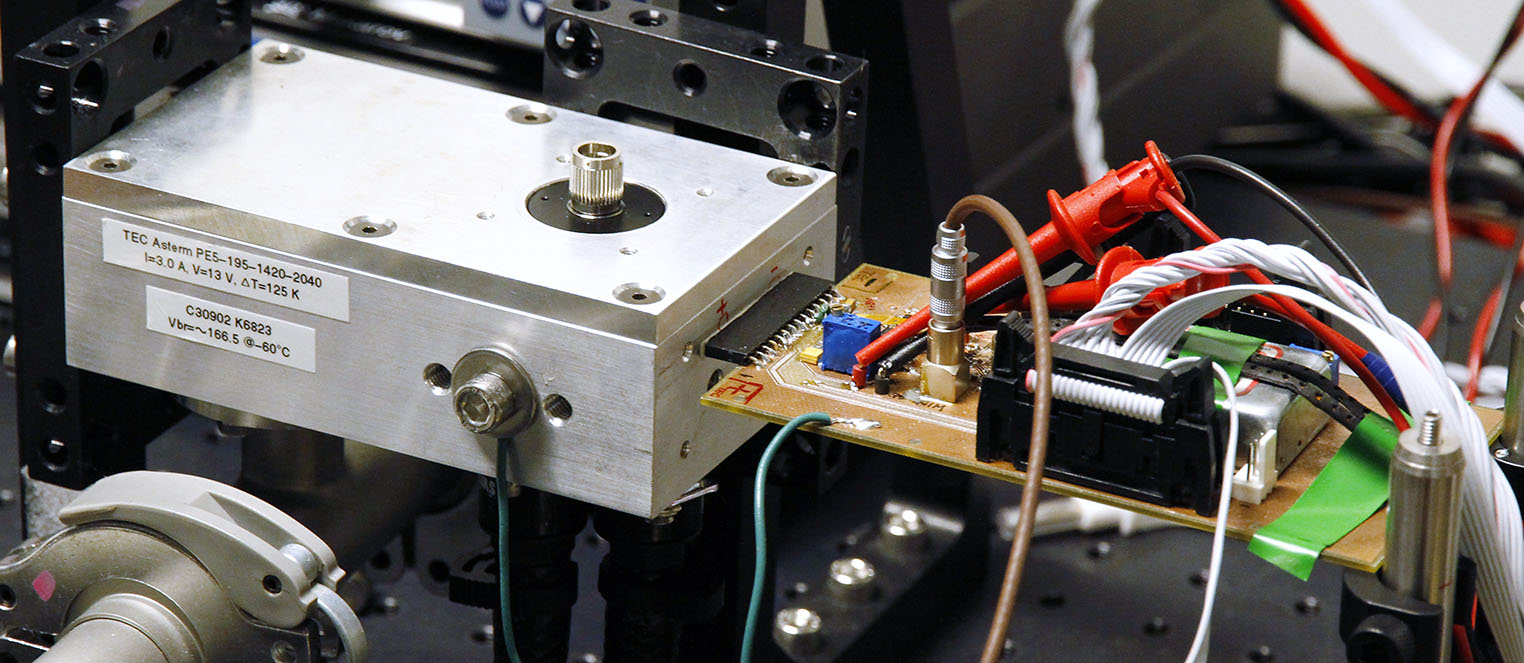}
  \caption{\label{fig:mech_design}Photo of the detector package. On the left is the vacuumed detector package with an optical window (covered with a black cap), and on the right is the electronics board with its metal shield removed. However the detector cannot be used without proper shielding of its PCB because of interference with outer sources, e.g.,\ mobile phones.}
\end{figure}

\begin{figure}
	\includegraphics[width=\columnwidth]{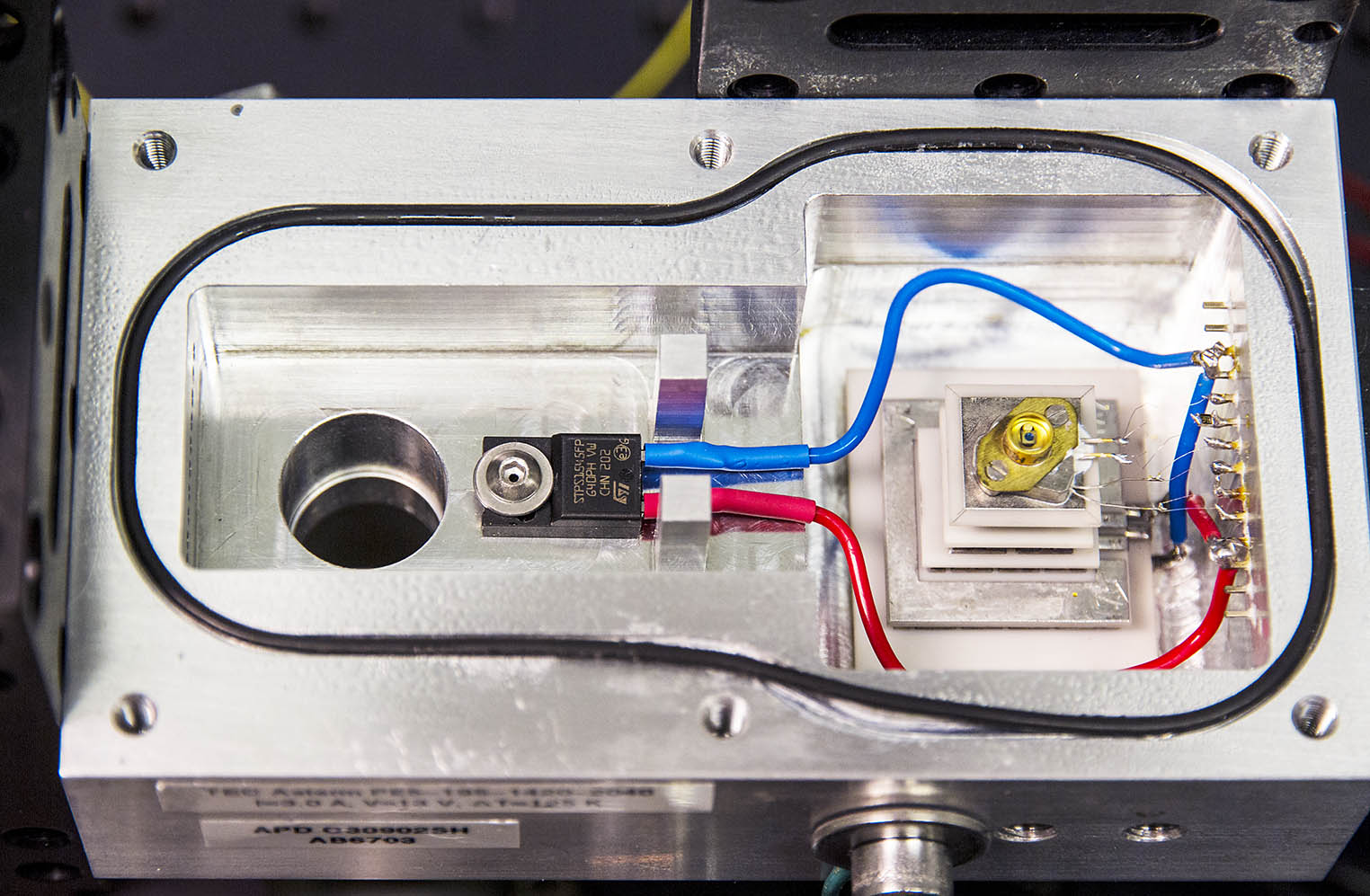}
  \caption{\label{fig:detector_open}Vacuum package with its lid removed. Inside the cavity, the round hole at left leads to a vacuum hose, and at right the APD in a Kovar holder is mounted atop the five-stage thermoelectric cooler. The APD is connected to electrical feed-throughs on the right-hand side wall via Pt wires.}
\end{figure}

The mechanical design is shown in \cref{fig:mech_design,fig:detector_open}. An aluminum alloy box is closed tightly with a lid sealed with a rubber O-ring and a vacuum lubricant. A five-stage thermo-electric cooler (TEC; Osterm PE5-195-1420-2040) is used to cool the APD placed in a holder on the cold plate of TEC. The holder is made of Kovar alloy to prevent destruction caused by a difference in thermal expansion coefficients between TEC ceramics and the holder material. To achieve temperatures about $-100~\celsius$ the package is evacuated to prevent convective heat transfer and also condensation. A vacuum turbo pump is always operating during the SPD operation, providing a vacuum level of $10^{-5}$~Torr, although $10^{-3}$~Torr already reduces convection sufficiently for thermal performance within $1~\celsius$ of the maximum possible.  All electrical connections to the cold plate are soldered via $50~\micro\meter$ diameter annealed Pt wires, to reduce heat conduction. The hot side of the TEC is cooled with $+14~\celsius$ water provided by a closed-loop chiller (ThermoTek T255P). 

The temperature of the APD is measured by a platinum resistance temperature detector (Omega RTD; TPT100KN1510), epoxied in the holder, and connected via a 4-wire scheme to eliminate errors caused by wire length differences. A temperature controller for the TEC is custom made in our lab, but instruments with similar parameters are available commercially. At the lowest achieved temperature of $-104~\celsius$ the TEC is running at its highest settings of 13~V and 3~A, consuming 39~W of electrical power.

\subsection{Electronics}
\label{sec:electronics}

Our new electronics design is an improvement on the previous version, used in Ref.~\onlinecite{kim2011}. We have implemented a simple and reliable passive quenching scheme with quenching resistance of $403~\kilo\ohm$, similar to one described in Ref.~\onlinecite{cova1996} as a passive quenching circuit with current-mode output. Its maximum detection rate of 0.2--0.4~Mcps is lower compared to active quenching circuits, but sufficient for applications with low signal rate, such as long distance free-space quantum communications that require very low dark counts level. The long dead time (${>}1~\micro\second$) is not a problem for the low-signal-rate application, and furthermore, it suppresses afterpulses.

In our new design we have placed the TEC controller and signal detection-and-conditioning circuits on separate printed circuit boards, in order to avoid electrical cross-talk interference that we observed sometimes with our previous design. Also, in our new circuit we use a faster comparator (Analog Devices ADCMP581) with adjustable threshold voltage, in an attempt to reduce jitter. Most measurements in this paper are taken at the optimum threshold setting of $40$--$60~\milli\volt$, unless specified otherwise. The detection circuit has transistor-transistor logic (TTL) and nuclear instrumental mode (NIM) outputs.

A $0$--$500~\volt$ high-voltage bias supply (EMCO CA05P) is used. Our circuit implements optional remote diagnostic and control of the detector parameters, for future use of this SPD in various experiments.

\section{SPD Characterization procedure}
\label{sec:characterization}

\begin{figure}
  \includegraphics[width=\columnwidth]{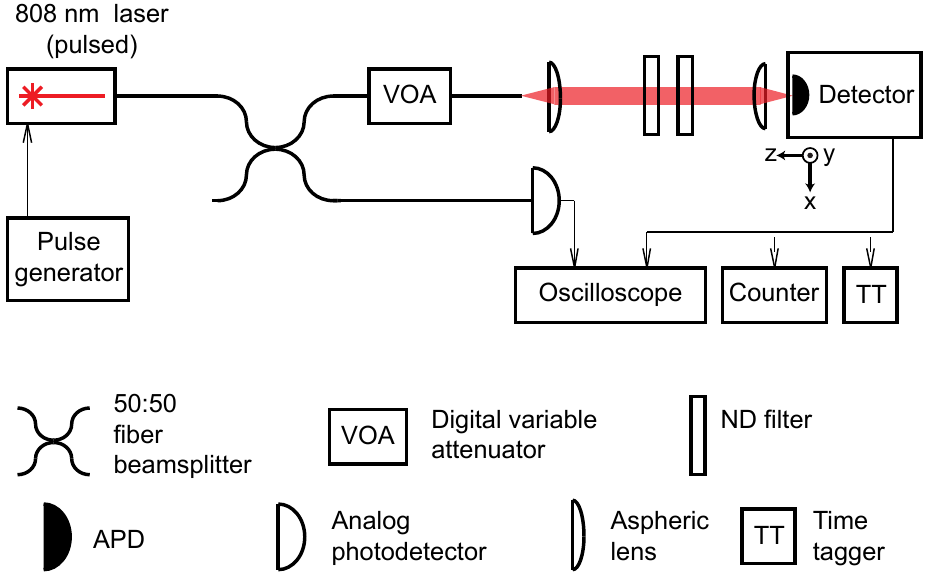}
  \caption{\label{fig:jitter_scheme}Characterization setup. For DCR and detection efficiency measurements the output of the SPD is connected to the counter. For afterpulsing analysis the SPD output is connected to a time stamp unit (time tagger; TT). For breakdown voltage and timing jitter measurements the SPD is connected to the oscilloscope. XY translation stage allows to scan photosensitive area of the SPD.} 
\end{figure}

For characterization of our detector we use a scheme shown in \cref{fig:jitter_scheme}. First, the APD's breakdown voltage is determined, then DCR is measured with a detector lid in place and the laser switched off. Then, the detection jitter, detection efficiency, and afterpulsing probability are measured.

\textsl{\textbf{Breakdown voltage}} is determined by an extrapolation method \cite{haitz1964}. After initially finding a bias voltage for single-photon operation roughly $20$--$30~\volt$ above breakdown, the bias voltage of the APD is gradually decreased. The avalanche amplitudes at different bias voltage values are recorded. About 10 points are measured until the bias voltage reaches the breakdown level. The results are plotted on a chart of the avalanche amplitude versus bias voltage, showing a trend that is mostly linear except for voltages near the breakdown. Then, the linear part of the resulting function is extrapolated and the breakdown voltage is determined as its intersection with the zero avalanche amplitude line. We have found that this method allows to determine the breakdown voltage with better than $\pm 0.5~\volt$ precision.

\textsl{\textbf{Dark count rate.}} During DCR measurement the detector's optical window is covered with a screw-in metal cap (\cref{fig:mech_design}) and room lights are kept off to ensure complete blackout. Counts are averaged over $100~\second$ using a counter (Stanford Research Systems SR620) to minimize uncertainty.

Besides the effects contributing to DCR mentioned in \cref{sec:intro}, an additional contribution comes from black body radiation, when photons emitted by the detector package walls into to the inner cavity are detected. We first theoretically estimate this contribution, assuming the inner cavity of the SPD to be a black-body at room temperature. The mean number of photons shorter than $900~\nano\meter$ impinging on the APD photosensitive area is calculated to be less than one photon per hour. This effect can be neglected, because our experimentally measured DCR is several orders of magnitude higher than that.

\textsl{\textbf{Detection efficiency}} is measured using a $808~\nano\meter$ pulsed laser (\cref{fig:jitter_scheme}) firing at 30~kHz repetition rate. The laser pulses are attenuated down to 56500~photons per second, using neutral density filters and digital attenuators calibrated at $808~\nano\meter$. Detection efficiency is calculated as a ratio of detected $N_{\text{det}}$ to expected $N_{\text{sent}}$ photons
	\[\eta=\frac{N_{\text{det}}-\text{DCR}}{N_{\text{sent}}},
\]
where $N_{\text{sent}}$ is determined as 
	\[N_{\text{sent}}=\frac{P\lambda}{fhc},
\]
where $P$ is power of the laser measured by a power meter before the calibrated attenuators and calculated to the detector point, $\lambda$ is the wavelength, $f$ is the laser pulse rate, $h$ is Planck's constant, and $c$ is the speed of light. Unfortunately, this method is not very precise, because of difficulty of high-precision calibration of optical components and the power meter, resulting in total error of several percent, e.g.,\ $\pm10\%$ in \cite{kim2011, bugge2014}. If a better precision is required, a three-attenuator method \cite{marsili2013} can be implemented.

\textsl{\textbf{Afterpulsing probability}} is calculated from  $10^6$ dark counts recorded using a time tagger (TT; UQD Logic16), with resolution of $78.125~\pico\second$. The recorded data is processed according to our method described in \cref{sec:afterpulsing}. Some results are calculated from data concatenated over several discontinuous sessions, because of long time necessary for data acquisition at low temperatures (5.5~days at $-100~\celsius$).

\textsl{\textbf{Detection timing jitter}} is measured using an oscilloscope (4~GHz bandwidth LeCroy 640Zi) in a histogram mode. Bright laser pulses from 808~nm laser (see \cref{fig:jitter_scheme}) are divided into two arms; one connected through a linear photodetector to the oscilloscope and the other attenuated below single photon level and focused to $25~\micro\meter$ spot at the SPD photosensitive area. The APD's avalanche signals are connected to another oscilloscope's input. Then we build a histogram of time delays between the laser pulses and the SPD output over $10^6$ samples, and determine timing jitter of the SPD as a full width at half maximum (FWHM) of this histogram. An example of the resulting histogram is shown in the inset in \cref{fig:jitter_vs_bias}.

Using an XY translation stage, we measure the dependence of the timing jitter on the position of the focused beam withing the APD's sensitive area.

\section{Long-time afterpulsing analysis}
\label{sec:afterpulsing}

\begin{figure}
  \includegraphics[width=\columnwidth]{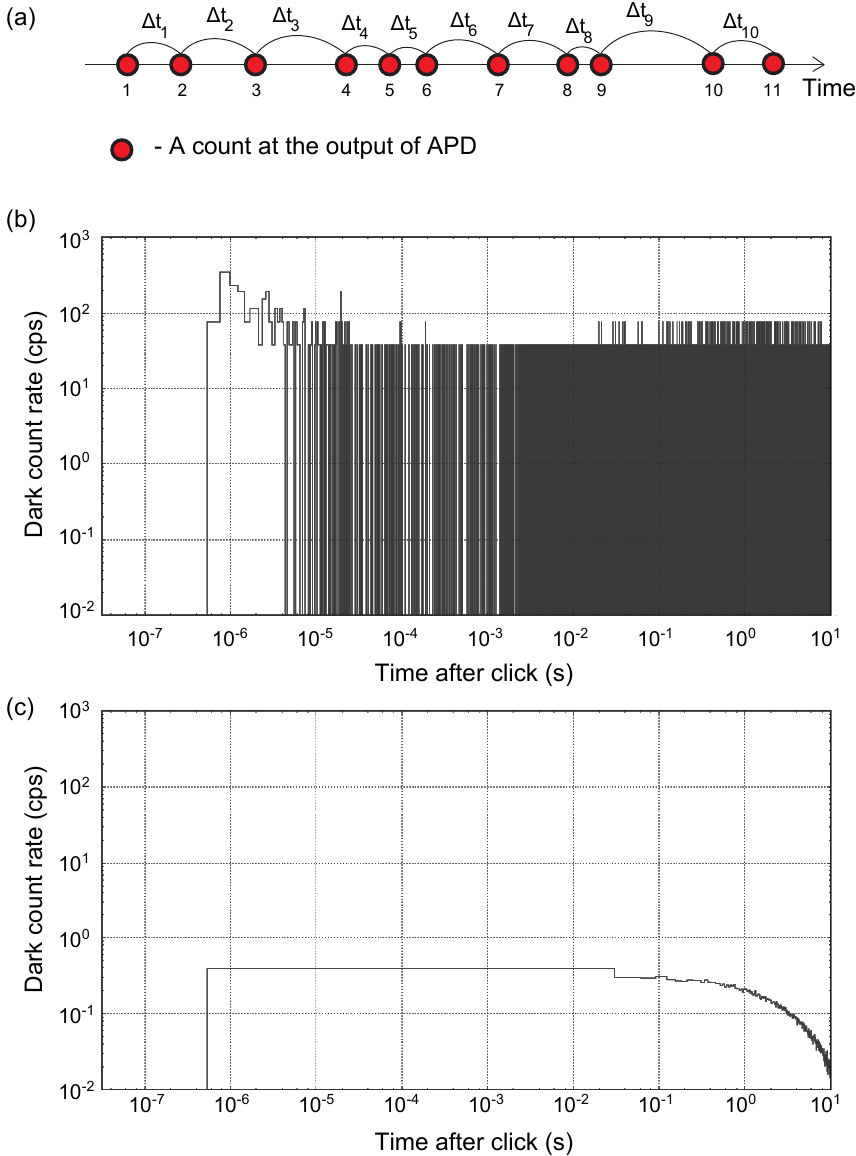}
  \caption{\label{fig:ap_analysis1}Standard afterpulsing analysis. (a)~Scheme for afterpulsing analysis that considers only time intervals between adjacent counts. Resulting histograms with (b) $228~\nano\second$ bins and (c)~$0.03~\second$ bins fail to show the entire afterpulse characteristic. Data size is 114109 counts. The dark counts were obtained from C30902SH APD at $-100~\celsius$ and $14~\volt$ over breakdown. The beginning of the first bin in (c) is artificially set at the expected dead time measured separately with an oscilloscope.}
\end{figure}

A common way to calculate afterpulsing probability is to analyze time intervals between neighboring counts \cite{yoshizawa2002b, giudice2003, itzler2012, humer2015}. A statistical distribution of time intervals between the adjacent counts is then computed and histogrammed. This method works reasonably well for short afterpulsing times and large quantities of data. However it falls apart when these conditions are not met. We give an example of this afterpulsing analysis from our Si APD at $-100~\celsius$. \cref{fig:ap_analysis1} illustrates the method of analysis (a) and presents two histograms (b, c), obtained by distribution of the analyzed time intervals in equally-sized bins. The first histogram is built using smaller bins of $228~\nano\second$ that provide enough resolution for the peak of the histogram representing afterpulses. However, this bin size is too small to correctly show the tail of the histogram that should represent dark counts. The second histogram is built using bigger bins of $0.03~\second$, and the tail of the histogram is now represented reasonably well, demonstrating an exponential decay caused by Poisson distribution of dark counts. However, $0.03~\second$ is longer than the afterpulsing time, and the peak of the histogram is lost. The dead time is determined from the first histogram (b) to be about $0.5~\micro\second$.

\begin{figure}
  \includegraphics[width=\columnwidth]{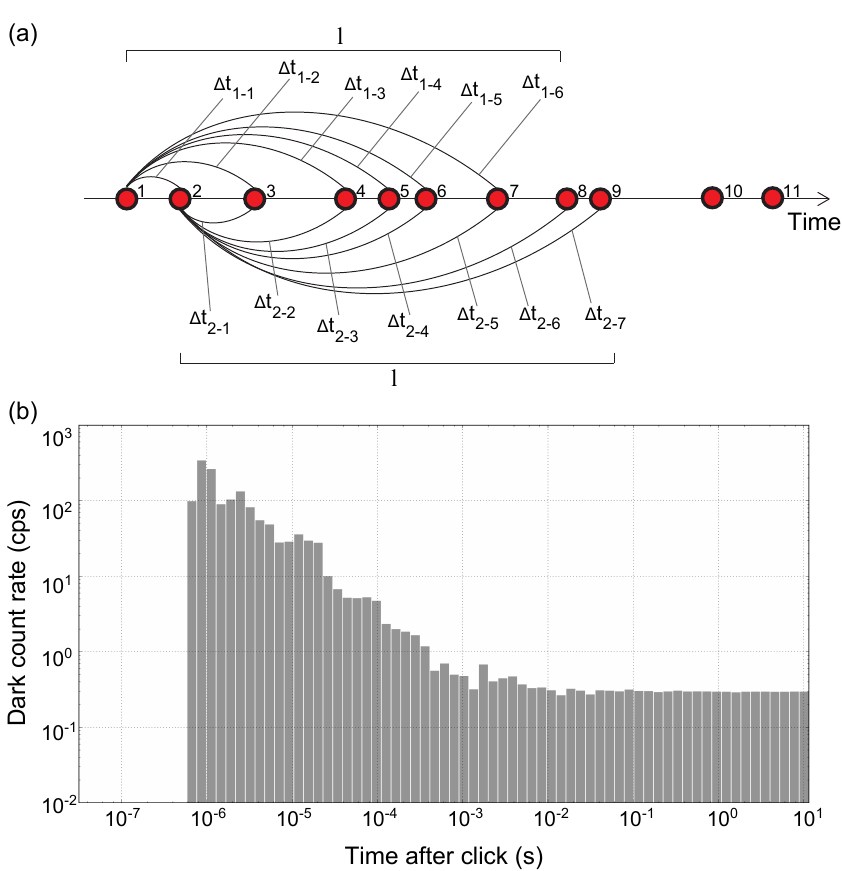}
  \caption{\label{fig:ap_analysis}Long-time afterpulsing analysis. (a)~Analysis scheme that considers all events within the window length $l$. (b)~Resulting histogram with an exponentially increasing bin size. This histogram uses the same data as \cref{fig:ap_analysis1} (114109 counts, C30902SH APD at $-100~\celsius$ and $14~\volt$ overvoltage) but visualises the entire afterpulse characteristic at once.}
\end{figure}

We develop an improved afterpulsing processing that allows to extract long-time afterpulsing time constants and amplitudes with better convergence. The main feature of our analysis is to include all time intervals between multiple subsequent counts during the time range $l$ being histogrammed. Another feature is the use of an exponentially increasing bin size that allows to simultaneously accurately visualise both the histogram's peak and tail.

We consider time intervals between a detection count [\#1 in \cref{fig:ap_analysis}(a)] and the all subsequent counts (\#\#2..7) during a certain time, up to $l=$10~s in the present example. The processing length $l$ should be chosen to exceed the longest possible afterpulsing time. The resulting time intervals $\Delta{t}_{1-1}..\Delta{t}_{1-6}$ are histogrammed. The procedure is then repeated starting from the next count \#2 and resulting in a set of time intervals $\Delta{t}_{2-1}..\Delta{t}_{2-7}$, then starting from the next count \#3 and so on until the end of the data is reached. 

The resulting histogram shown in \cref{fig:ap_analysis} (b) contains 128 bins whose size exponentially increases by a factor of 1.2, starting from $78.125~\pico\second$. This allows to have higher resolution on the left part of the histogram, and lower resolution on its right. This histogram features almost noiseless tail on the right and quite smooth curve of the peak on the left that represents afterpulses. From the histogram we can determine DCR, APD's dead time, and recharge time. The dead time starts after a photon detection (at 0~s), and lasts until counts appear again (at about $0.5~\micro\second$). The recharge time starts from the end of the dead time and lasts roughly until the peak value of the count rate (at about $0.3~\micro\second$). The plot levels off on the right to the DCR. The afterpulsing probability is calculated as the area of the histogram above the DCR level. Life time constants of trapped carriers $\tau$ can be found by fitting the decaying slope of the peak \cite{cova1991} a sum of exponents
\[P(t)=D+A_1\cdot e^{-t/\tau_1}+A_2\cdot e^{-t/\tau_2}+... ,\]
where $P(t)$ is a carrier emission probability, $D$ is DCR due to thermally generated carriers, $A_1$, $A_2$,~... are amplitudes of the different exponential components, and $\tau_1$, $\tau_2$,~... are life time constants of the trapped carriers. Our software implements fitting for up to four exponential components. Only lifetimes longer than the detector dead time (about $0.8~\micro\second$) for our passive quenching circuit can be determined.

\section{Results}
\label{sec:results}

A sample APD C30902SH (K6823) was cooled down and fully characterized at several temperatures in $-100$ to $0~\celsius$ range at $14~\volt$ above its breakdown voltage. DCR was measured at $7$, $14$, $28$ and $40~\volt$ above breakdown voltage in temperature range from $-104$ to $-30~\celsius$. Detection timing jitter was measured from $-60$ to $-30~\celsius$ and at several bias voltages.

\begin{figure}
  \includegraphics{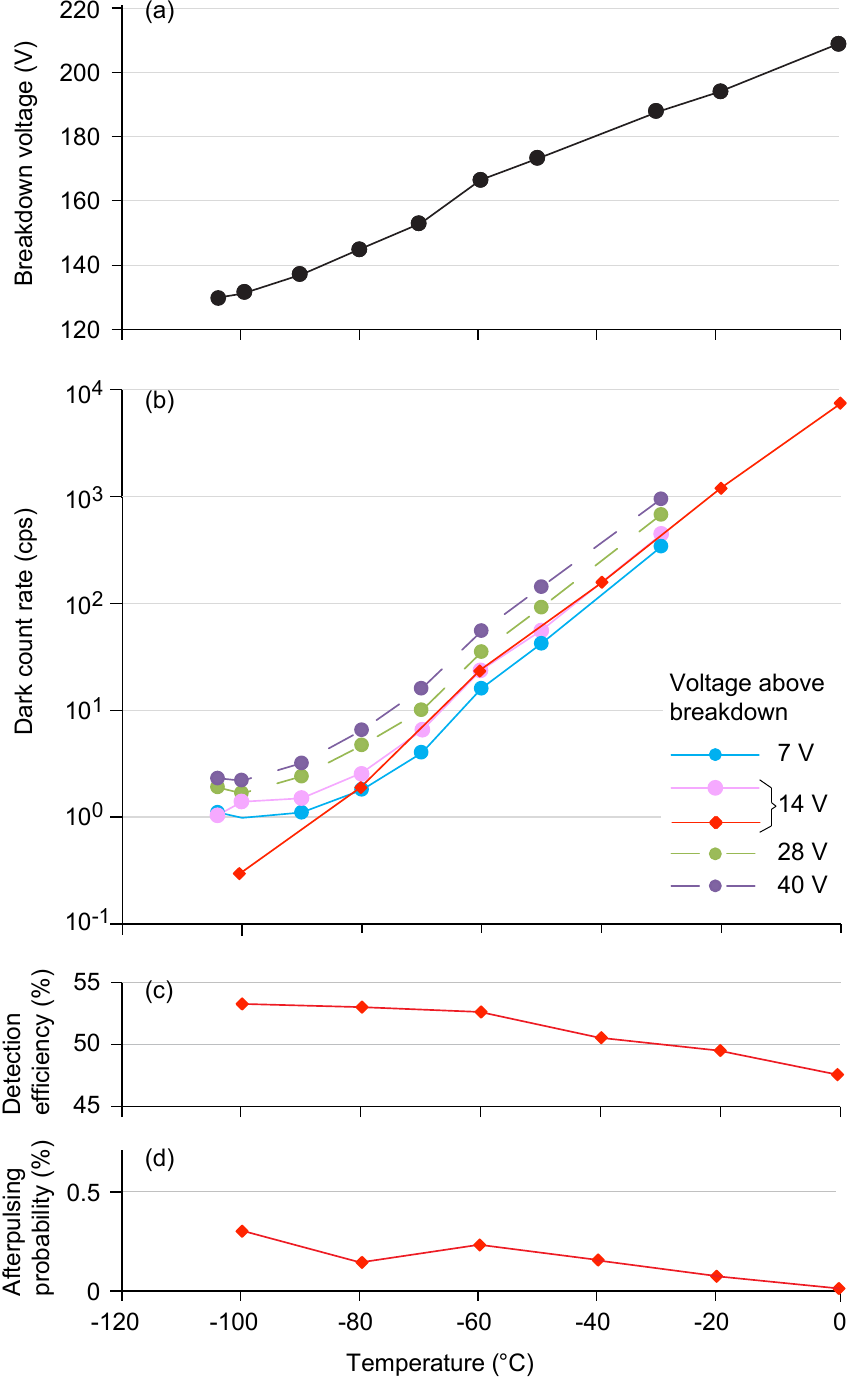}
  \caption{\label{fig:characterization}Detector characteristics of C30902SH: (a) APD breakdown voltage, (b) DCR, (c) detection efficiency, and (d) afterpulsing probability. The latter two were measured at 14~V over breakdown voltage. The data points denoted by red diamonds were measured several months later than the rest of the data. }
\end{figure}

The breakdown voltage [\cref{fig:characterization}(a)] increases with temperature about linearly with a coefficient of $0.8~\volt {\celsius}^{-1}$. This is a typical behavior of Si APDs \cite{endnote1}, which as we show here extends down to $-104~\celsius$.

DCR as a function of temperature is shown in  \cref{fig:characterization}(b). The lowest achieved DCR of 0.3~cps was observed for the APD biased 14~V and cooled down to $-100~\celsius$. There was a discrepancy between DCR measurements done at different times. The four curves in \cref{fig:characterization}(b) with dots were measured at one time, and the curve with diamonds for $14~\volt$ over breakdown voltage was measured several months later during collecting data for afterpulsing analysis. Down to $-70~\celsius$ the curves match perfectly, but then one curve levels off whereas the other continues linearly. It could be due to a poor black-out during the measurement. Another possible explanation could be a ``memory effect'': after a strong illumination an APD has a higher DCR for a long time up to $24~\hour$ \cite{c309xxsh-excelitas,spcmaqrh-excelitas}.

To verify experimentally the contribution of black body radiation to our DCR measurement, we performed DCR measurement with the detector lid cooled down below $0~\celsius$, and compared it with measurement when the lid was at room temperature. No notable change in DCR was registered.

Detection efficiency varies in the range 48 to 53\% [\cref{fig:characterization}(c)], decreasing slightly at higher temperatures. However this apparent decrease of efficiency can be partially explained by high DCR leading to saturation of the detector.

\begin{figure}
  \includegraphics{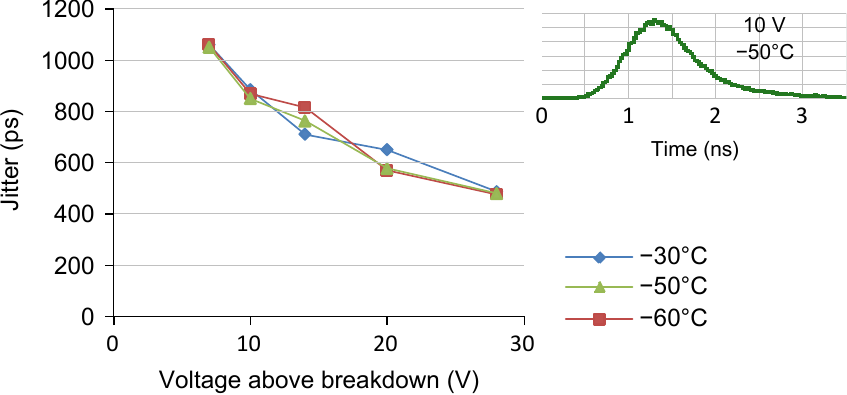}
  \caption{\label{fig:jitter_vs_bias}Detection timing jitter FWHM as a function of the bias voltage measured for C30902SH at three different temperatures: $-30$, $-50$, $-60~\celsius$. As the applied voltage increases, the jitter decreases. An example of a histogram is shown in the \textbf{inset}, measured on $10^5$ samples at the following conditions: $-50~\celsius$, comparator threshold set at $50~\milli\volt$, bias voltage 10 V above breakdown. Its FWHM is $850~\pico\second$.}
\end{figure}

We have measured detection timing jitter of C30902SH depending on its bias voltage, temperature, comparator voltage level and position of the beam at the photosensitive area. 
The timing jitter decreases with the rise of APD's bias voltage in the same way for all three measured temperatures (\cref{fig:jitter_vs_bias}). This happens due to increase of avalanche propagation speed \cite{cova1996, haitz1964, lacaita1993a, lacaita1993b, cova1987, cova1989, spinelli1997}. An example of the jitter distribution is shown in the inset in \cref{fig:jitter_vs_bias}.

\begin{figure}
	\includegraphics{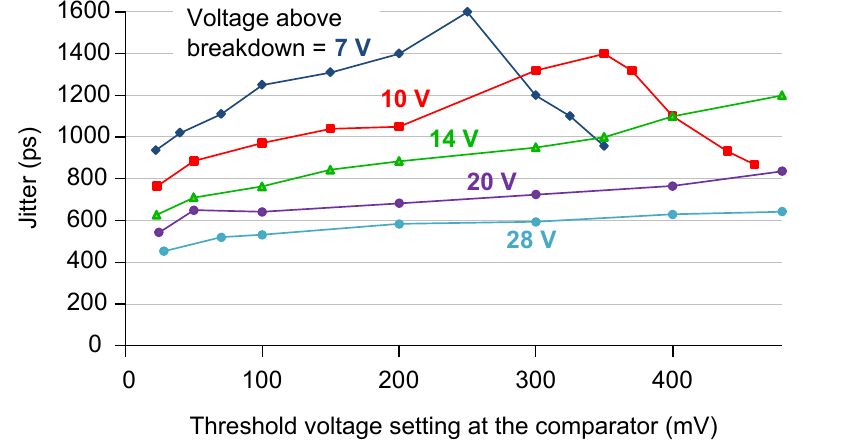}
	\caption{\label{fig:jitter_vs_treshold}Detection timing jitter FWHM as a function of comparator threshold voltage for C30902SH at $-30~\celsius$. Data taken at $-50$ and $-60~\celsius$ are very similar.}
\end{figure}

Dependence of the timing jitter of the APD on the comparator setting is presented in \cref{fig:jitter_vs_treshold}. Decrease of the comparator threshold voltage in the avalanche registration circuit leads to a decrease of the jitter.

Avalanches in a Si APD have a randomly varying amplitude and slew rate, as can be observed with an oscilloscope. The large-amplitude avalanches always have higher slew rate, therefore they cross the comparator threshold earlier and thus get registered earlier. To minimize the time scatter between the registration of the avalanches of different amplitude, the comparator threshold level should be set as low as possible. However, it is in practice limited from below by electronic noise and cross talk. Another possible solution could be the use of a fractional comparator, instead of the constant level comparator that it implemented in our circuit.

We have checked the dependence of the APD timing jitter on the position of incident light within the sensitive area of the APD. The measurement was done at $-50~\celsius$, at five different bias voltages, same as in \cref{fig:jitter_vs_bias}. The beam was focused into a spot of $25~\micro\meter$ in diameter. The results demonstrate notably lower jitter at the center, with up to $250~\pico\second$ difference comparing to the edge. The distance between the center and the edge beam positions was $237~\micro\meter$. The data represented in \cref{fig:jitter_vs_bias} and \cref{fig:jitter_vs_treshold} were measured with the beam focused at the center of APD's sensitive area.

We have calculated afterpulsing probability for C30902SH APD using our method (\cref{sec:afterpulsing}). The resulting temperature dependence is shown in \cref{fig:characterization}(d). Afterpulsing notably increases with cooling, but does not exceed 0.35\% at the lowest tested temperature of $-100~\celsius$.

\begin{table}
\caption{\label{table:life_times}Trap fitting parameters and afterpulsing histograms at six temperatures. \textbf{\textit{D}} denotes  thermally generated (constant) dark count level. The fit given by {\textit{${\tau}_i$}}, {\textit{$A_i$}}, and \textbf{\textit{D}} is plotted as a solid line.}
\newlength{\colwidtht}
\newlength{\colwidthtime}
\newlength{\colwidthampl}
\newlength{\colwidthfig}
\setlength{\colwidtht}{0.8cm}
\setlength{\colwidthtime}{1.5cm}
\setlength{\colwidthampl}{1cm}
\setlength{\colwidthfig}{4.7cm}
\begin{tabular}[t]{c c c c}
\hline\hline
\makecell{\bm{$T$}\\ \textbf{(\bm{$\celsius$})}} & \makecell{\bm{$\tau$}\\ \textbf{(\bm{$\second$})}} & \makecell{\bm{$A$}\\ \textbf{(cps)}} & \makecell{\textbf{Histogram}} \\
\hline
\parbox[t]{\colwidtht}{{~} \\ \hfill 0} & \parbox[t]{\colwidthtime}{{~}\\ \hfill {\textit{\textbf{D}}}} & \parbox[t]{\colwidthampl}{{~}\\ \hfill 7603} & {\raisebox{-\totalheight}{\includegraphics[width=\colwidthfig]{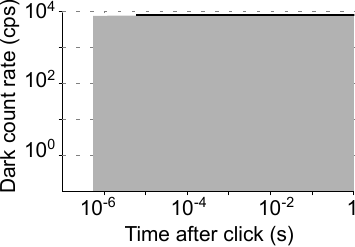}}} \\
\parbox[t]{\colwidtht}{{~}\\ \hfill $-20$} & \parbox[t]{\colwidthtime}{{~}\\ \hfill $1.10\cdot 10^{-5}$ \\ \hfill {\textit{\textbf{D}}}} &  \parbox[t] {\colwidthampl}{ {~}\\ \hfill 60 \\ \hfill 1212} & \raisebox{-\totalheight}{\includegraphics[width=\colwidthfig]{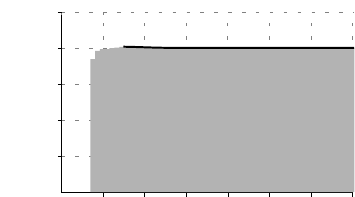}} \\
\parbox[t]{\colwidtht}{{~}\\ \hfill $-40$} & \parbox[t]{\colwidthtime}{{~}\\ \hfill $3.73\cdot 10^{-6}$ \\ \hfill $1.72\cdot 10^{-5}$ \\ \hfill $1.90\cdot 10^{-4}$ \\ \hfill{\textit{\textbf{D}}}} & \parbox[t]{\colwidthampl}{{~} \\ \hfill 108 \\ \hfill 49 \\ \hfill 5 \\ \hfill 155} & \raisebox{-\totalheight}{\includegraphics[width=\colwidthfig]{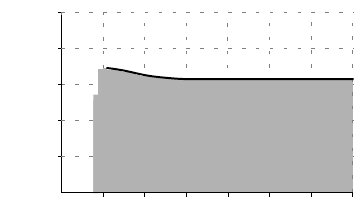}} \\
\parbox[t]{\colwidtht}{{~}\\ \hfill $-60$} & \parbox[t]{\colwidthtime}{{~}\\ \hfill $5.36\cdot 10^{-6}$ \\ \hfill $3.94\cdot 10^{-5}$ \\ \hfill $1.84\cdot 10^{-4}$ \\ \hfill{\textit{\textbf{D}}}} & \parbox[t]{\colwidthampl}{{~}\\ \hfill 233 \\ \hfill 18.5 \\ \hfill 3 \\ \hfill 24} & \raisebox{-\totalheight}{\includegraphics[width=\colwidthfig]{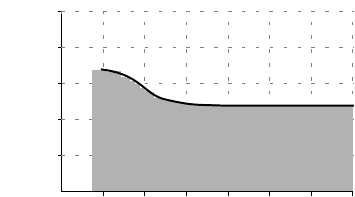}} \\
\parbox[t]{\colwidtht}{{~}\\ \hfill $-80$} & \parbox[t]{\colwidthtime}{{~}\\ \hfill $1.37\cdot 10^{-6}$ \\ \hfill $6.13\cdot 10^{-6}$ \\ \hfill $2.44\cdot 10^{-5}$ \\ \hfill $2.06\cdot 10^{-4}$ \\ \hfill{\textit{\textbf{D}}}} & \parbox[t]{\colwidthampl}{{~}\\ \hfill 133 \\ \hfill 61 \\ \hfill 22 \\ \hfill 2.5 \\ \hfill 1.9} & \raisebox{-\totalheight}{\includegraphics[width=\colwidthfig]{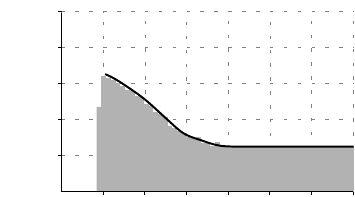}} \\
\parbox[t]{\colwidtht}{{~}\\ \hfill $-100$} & \parbox[t]{\colwidthtime}{{~}\\ \hfill $3.6\cdot 10^{-7}$ \\ \hfill $3.4\cdot 10^{-6}$ \\ \hfill $2.75\cdot 10^{-5}$ \\ \hfill $4.82\cdot 10^{-4}$ \\ \hfill{\textit{\textbf{D}}}} & \parbox[t]{\colwidthampl}{ {~}\\ \hfill 3446 \\ \hfill 123 \\ \hfill 37 \\ \hfill 2.3 \\ \hfill 0.3} & \raisebox{-\totalheight}{\includegraphics[width=\colwidthfig]{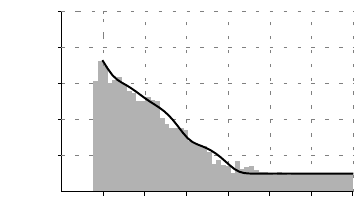}} \\
 & & \\
\hline\hline
\end{tabular}
\end{table}

Results of our attempted calculation of trap lifetimes are presented in \cref{table:life_times}. In order to exclude dead-time effects, the fitting starts from the first bin at or after the global peak that is followed by four more bins with monotonically decreasing values. (For example, in \cref{fig:ap_analysis}(b) the fitting would start from the 6th non-zero bin counting from the left.) For $0~\celsius$ there is no afterpulsing peak, accordingly, no trap lifetime constants are determined. The decay slope at $-20~\celsius$ is approximated with one exponent, at $-40$ and $-60~\celsius$ with three exponents, at $-80$ and $-100~\celsius$ with four exponents. At $-20~\celsius$ the afterpulsing peak is hardly noticeable and noisy; as a result, the fitting curve starts at a later point than at lower temperatures. We reach a good fit at $-40$ to $-80~\celsius$. However at $-100~\celsius$, where we use the sum of four exponents, the fit is not perfect. We tried to use more than four exponents in this case, but it did not improve the fit. The estimated trap lifetime constants at this temperature have the widest range, between $0.36$ and $482~\micro\second$. Peculiarly, the shortest life time is shorter than the dead time. The data at $-100~\celsius$ is somewhat noisy, owing to the very low DCR and limited measurement time (5.5~days).

Unfortunately we had time to characterize fully our detector with only one APD sample. One more sample of Excelitas C30902SH L0622 was tested for DCR at temperatures down to $-90~\celsius$ and demonstrated a similar level of DCR (0.58~cps at $-91~\celsius$).

We remark that the methodology introduced in this paper has subsequently been used to characterize many more APD samples in Refs.~\onlinecite{anisimova2017a,lim2017}. That testing included three other APD models: Excelitas C30921SH and SLiK, and Laser Components SAP500S2. The afterpulse characterization methodology has also been further refined in Ref.~\onlinecite{lim2017}, where periodic weak laser pulses are applied to the APD at repetition rate $< 1/l$. This increases the count rate without affecting the afterpulse distribution, and allows to collect data faster at low temperatures.

\section{Applications}
\label{sec:modeling}

Our new detector demonstrates a very low DCR without decreased performance in other important parameters (detection timing jitter, detection efficiency, afterpulsing). Having such SPDs can be beneficial for quantum communication over high-loss channels. We illustrate this with a numerical simulation considering a quantum key distribution experiment via a satellite in a geostationary orbit (GEO). Since GEO satellites stay on a fixed point in the sky, the demands for tracking technology can be relaxed, and uninterrupted quantum connections can be sustained over many hours. Such an experiment has not been realized as of today, which is in part due to insufficient detector technology. With the SPDs presented in this work, even a dual downlink from a GEO satellite could be carried out. Such a scenario has the additional advantage that the satellite does not need to be trusted for quantum communication \cite{yin2020}.

\begin{figure}
  \includegraphics{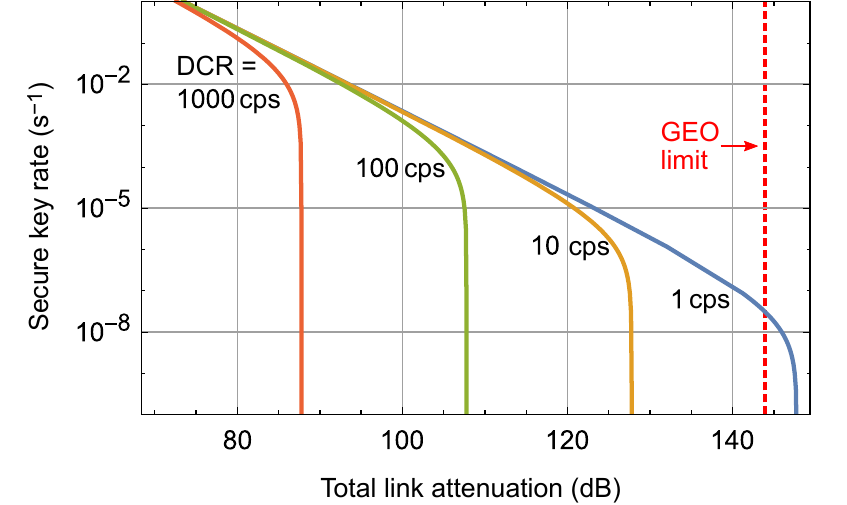}
  \caption{\label{fig:simulation} Simulation results for a dual-downlink quantum key distribution experiment via satellite for different DCR levels of both Alice's and Bob's SPDs.} 
\end{figure}

The experimental setup in consideration deploys a source of polarization entangled photons on the GEO satellite. These photons are distributed via two free-space downlink channels to two separate ground stations. The single-channel attenuation over such a link has been experimentally specified as about $69~\deci\bel$ \cite{gunthner2017}. Considering $50\%$ detector efficiency, this amounts to a total dual-link loss of $144~\deci\bel$. At each ground station, the polarization state of the photons is measured using two detectors of the design presented here. From these measurements, an unconditionally secure key can be created between the two ground stations due to the photons' quantum correlations \cite{bennett1992a}. \Cref{fig:simulation} shows the expected key rates for different DCR scenarios, following calculations devised in \cite{ma2007}. We assumed a perfect source at the satellite with a pair creation rate of 50~Mcps and a detection timing window of $1~\nano\second$ (to accommodate the detector jitter). It can clearly be seen that the low-DCR SPDs shown in this work would be an enabling technology for GEO links. Assuming 1~cps DCR, losses as high as $148~\deci\bel$ can be tolerated for creation of a secure key. This beneficial effect of low DCR can be understood as follows: In order to correctly identify photon correlations between the two ground stations, each detection event is recorded in time, and only matching events are considered as pairs. In a regime of very high loss, only a tiny fraction of the actually quantum signal arrives at the detectors, in our case about 1~cps per ground station. If the DCR is of about the same order of magnitude, it is likely for a photon to be mistakenly correlated with a dark count on the other station instead of its actual partner, which degrades the signal-to-noise ratio (SNR). If the SNR drops below 8.1, no secure key can be generated. With our detectors, the tolerable loss for the SNR limit can be substantially increased such that dual-downlink GEO quantum key distribution would be made possible.

In addition to the GEO scenario just described, the  teleportation experiment on the Canary Islands over $143~\kilo\meter$ atmospheric link \cite{ma2012} illustrates the importance of detector performance. High amount of airborne sand, rain, fog, and even snow can significantly decrease transparency of the atmosphere there, which indeed happened on the first attempt of teleportation in the summer of 2011. The link attenuation was on the order of $35~\deci\bel$ and no usable data could be obtained with standard commercial SPDs in Bob. In the aftermath they were replaced with a previous version of our SPDs, which had DCR of about 15--20~cps and the same sensitive area of $0.5~\milli\meter$ diameter \cite{kim2011,ma2012}. A lucky weather condition on the second attempt in April 2012 provided a clear atmosphere resulting in about $30~\deci\bel$ channel loss. However, according to our simulations, the lower-noise SPDs could have saved this experiment even if the higher $35~\deci\bel$ loss recurred. Our last generation of the SPDs presented in this paper demonstrates even lower DCR that makes them good candidates for use in long-distance quantum communication experiments.

\section{Discussion and conclusion} 
\label{sec:discussion}

We have built and characterized a very-low-noise Si-APD based SPD, in a custom compact package and cooled to  $-100~\celsius$. All main parameters (except the maximum count rate) of our SPD are in a suitable range for use in long-distance quantum communication experiments: DCR is below 1~cps, afterpulsing at the lowest temperature does not exceed 0.35\%, detection efficiency is about 50\%, detection timing jitter varies between $500$ and $1050~\pico\second$ depending on of the APD bias voltage. Using SPDs with such parameters could be beneficial for experiments of quantum communications over high-loss channels. Afterpulses can be further reduced by discarding them in post-processing, depending on application requirements \cite{yoshizawa2002b, yoshizawa2003}.

The combination of parameters demonstrated in our SPD is not available in commercial products. For example, ID Quantique ID100VIS \cite{0-id100vis} detector module has a similarly low DCR, however its photosensitive area is 600 times smaller and detection efficiency peaks at $35\%$ at $500~\nano\meter$. ID120VIS \cite{2-id120vis} has the same photosensitive area as ours and a slightly higher detection efficiency, at the cost of a much higher DCR of $\lesssim200$~cps.

For measuring afterpulsing probability, we have developed an improved method of analysis for long trap lifetimes. Our algorithm can be implemented with minor adjustments for analysis of data collected from an APD illuminated with weak periodic light pulses \cite{lim2017}. Furthermore, we have implemented a curve fitting procedure to our data to calculate lifetime constants for carrier traps, and their corresponding amplitudes. Finally, we have illustrated that the low-noise SPD is an enabling technology for a dual-downlink GEO satellite QKD and a $143~\kilo\meter$ terrestrial teleportation experiment. 

The results of the present research have been used for planning detector design for a future space mission \cite{pugh2017} and for finding a way of mitigating radiation damage in APDs \cite{anisimova2017a, lim2017}.

\bigskip

\noindent \textbf{Acknowledgments}\\
We thank C.~Kurtsiefer, Y.-S.~Kim, R.~Romero and I.~Radchenko for contributions to electronics and mechanical design, R.~Blach and A.~Dube for CNC machining, I.~DSouza and X.-s.~Ma for discussions. 

\medskip

\noindent \textbf{Funding}\\
This work was supported by European Space Agency, Industry Canada, NSERC (programs Discovery and CryptoWorks21), CFI, Ontario MRI, and U.S.\ Office of Naval Research.

\medskip

\noindent \textbf{Availability of data and materials}\\
Raw experimental data and calculations can be obtained from the corresponding author upon a reasonable request.

\medskip

\noindent \textbf{Competing interests}\\
The authors declare that they have no competing interests.

\medskip

\noindent \textbf{Authors' contributions}\\
E.A.\ designed the detector package, conducted the experiment, analysed data, and wrote the paper with contributions from all authors. S.S.H.\ and D.N.\ programmed the analysis software and analysed data. M.B.\ and D.N.\ conducted measurements. S.N.\ calculated the GEO satellite performance. V.M.\ designed the detector electronics. V.M.,\ T.J., and R.U.\ supervised the study.

\end{document}